\documentclass[aps,final,notitlepage,oneside,twocolumn,nobibnotes,nofootinbib,
superscriptaddress,noshowpacs,centertags,showkeys]{revtex4-1}

\usepackage[english]{babel}
\usepackage{graphicx}
\usepackage{latexsym}
\usepackage{amssymb}
\usepackage{amsmath}
\usepackage{float}
\usepackage{color}

\begin{document}

\title{A planet or primordial black hole in the outer region of the Solar system and the dust flow near Earth’s orbit}
\author{Yu.N. Eroshenko}\thanks{e-mail: eroshenko@inr.ac.ru}
\affiliation{Institute for Nuclear Research RAS, Moscow, Russia}
\author{E.A. Popova}
\affiliation{Institute for Nuclear Research RAS, Moscow, Russia}
\affiliation{Pulkovo Observatory RAS, St. Petersburg, Russia}

\date{\today}

\begin{abstract}
In recent years, evidence has been obtained that in the outer region of the Solar System (in the inner part of the Oort cloud), at a distance  $\sim300-700$~AU from the Sun, there may be a captured planet or a primordial black hole. In this paper, we show that the gravitational scattering of dust particles in the same region on this object can transfer them to new elongated orbits reaching the Earth's orbit. With the mass of the captured object of the order of 5-10 Earth masses, the calculated dust flow near the Earth $\sim0.1-3$~$\mu$g~m$^{-2}$~year$^{-1}$ is  comparable in order of magnitude with the observed flow. This effect gives a joint restriction on the parameters of the captured object and on the amount of dust in the Oort cloud.
\end{abstract}

\maketitle 


\section{Introduction}

The observation of correlated movements of trans-Neptunian objects has led to the hypothesis that at the distance 300-700~AU from the Sun there may be a 9th planet \cite{Batetal19} or a primordial black hole (PBH) \cite{SchUnw20} with a mass 5-15 Earth masses, which perturbs their movement (see \cite{Napetal21} for the critical discussion of this hypothesis). Evidence of the presence of the 9th planet with approximately the same orbital parameters was also obtained in the work \cite{LykIto23}. Although the option with the PBH looks more exotic compared to the planet, the existence of such objects in the Universe is quite likely, and the probability of capturing the free-flying planets in the Galaxy and the PBH into the Solar System, as shown in \cite{SchUnw20}, is about the same. 

The possibility of the PBH formation in the early Universe was discovered by Ya.B.~Zeldovich and I.D.~Novikov in the theoretical work \cite{ZelNov66}, see reviews \cite{Dol18,CarSenYok20}. In recent years, PBHs have attracted increased attention, since they can explain part of the gravitational wave events of LIGO/Virgo \cite{Blietal16}. The mass spectrum of the PBH predicted in some models also corresponds well to the observations of \cite{DolPos20}. PBHs can explain the presence of quasars at large redshifts and the formation of early galaxies recently discovered by the J. Webb Space Telescope \cite{Guoetal23}. PBH remains one of the candidates for the role of dark matter (hidden mass) in the Universe. This possibility was discussed, in particular, in the work \cite{IvaNasNov94}. However, the fraction of PBHs in the composition of dark matter was strongly limited by various effects \cite{CarKuh20,GreKav20}. 

Next, for brevity, we will talk about PBH, but most of the reasoning applies to the captured planet. We will show that if there is an PBH on the periphery of the Solar System, then with its gravitational field perturbs the orbits of dust particles, and some of the particles can fall into the inner region of the Solar System, flying inside the Earth's orbit. If the mass of the PBH is of the order of $10M_\oplus$, and the total mass of dust in the inner Oort cloud has the same order of magnitude, then the dust flow calculated in this work turns out to be comparable to what is actually observed by counting dust particles on the Antarctica ice and by direct measurements on spacecraft. From these data, it was obtained that the flow of dust particles near the Earth's orbit is $\sim 3.0-5.6$~$\mu$g~m$^{-2}$~year$^{-1}$ \cite{Rojetal21}. Perturbations of orbits and ejections of large icy bodies by the field of the 9th planet have already been considered in \cite{BatBro21}, however, modelling was limited only to the outer region of the Solar System.

The interaction of PBH with the bodies of the Solar system has been studied in a number of works in various aspects. In the work \cite{SchUnw20}, the question of whether the PBH can play the role of the 9th planet, which perturbs the orbits of trans-Neptunian objects, is investigated. In the work \cite{SirLoe21}, an attempt was made to limit the frequency of passage through the Solar System of low-mass PBH $\sim10^{-10}M_\odot$ by the effect of increasing ellipticity of the orbits of trans-Neptunian objects, and it was shown that this effect is insignificant. For more massive PBHs (from the mass of the Earth to the mass of the Sun), in the work \cite{SirLoe21-2} restrictions based on the perturbation of the orbit of Neptune are obtained. The restrictions from seismic events triggered by the PBHs was considered in \cite{Luoetal12}.

In addition to dynamic constraints, limitations based on other effects can be found on the PBHs in the Solar System. In the work \cite{DokVol}, the presence of an PBH inside the Earth was excluded due to unacceptably large accretion and neutrino radiation.  The paper \cite{SirLoe20} discusses the constraints that can be obtained in the future by observing X-ray flashes. These flares should occur when the icy bodies on the periphery of the Solar System collide with the PBH. And in the work \cite{ArbAuf20} it is proposed to search for Hawking radiation of the radio frequency range using space micro-apparatus flying past the PBH.  In the work \cite{SchUnw20}, the constraint on annihilation radiation from the dark matter density spike forming around the PBH is considered. Such a constraint will occur if dark matter particles are capable of annihilation and if annihilation products are principally observable.

The new effect we are considering belongs to the class of dynamic effects. Based on it, it will also be possible to obtain constraints on the PBH in the Solar system after the structure and composition of the cloud of ice and dust at distances of 300-700~AU from the Sun is clarified. But already at the moment, it can be concluded that, with sufficiently plausible assumptions, the dust flow produced on Earth can be comparable in order of magnitude to the flow that is actually observed.

Detailed calculations of the scattering of light with the solar spectrum on particles of various compositions show that the ratio of the force of radiation pressure to the force of gravity of the Sun is $\leq3$~\% for particles of any composition with sizes $\geq10$~$\mu$m and non-metallic/non-graphite particles with sizes $\leq 0.02$~$\mu$m, whereas for of iron-containing substances and graphite, the specified ratio in the region of $\leq 0.02$~$\mu$m remains at the level of the order of one (see Fig.~7a and Fig.~7b in \cite{BurLamSot79}). Thus, our calculations are applicable to all dust or meteoroid particles larger than about 10~microns and to particles without iron and graphite content with dimensions less than 0.02~microns. Note that the class of particles we are considering intersects with the class of meteoroids (particles ranging in size from 30~$\mu$m to 1~m), so in this case, instead of dust, we can talk about meteoroids.  

This article is organized as follows. In the Section~\ref{tor} we construct a model distribution function for dust cloud particles. The Section~\ref{algsec} describes a numerical algorithm used to calculate the dust flow into the Earth's orbit. The Section~\ref{tejsec} provides a calculation of the lifetime of dust particles in orbits entering the Earth's orbit, taking into account their ejection by the gravitational field of Jupiter. The Section~\ref{resultsec} shows the final calculation results taking into account both effects: dust particle coming and ejection. The  Section~\ref{zaklsec} provides some conclusions. In the Appendix, for comparison with numerical calculation, simple estimates are given in order of magnitude for the flow of dust particles.


\section{The dust cloud model}
\label{tor}

According to observations, interstellar gas-dust clouds in the Galaxy consist of solid dust particles by about 1\% by mass. If we take into account that all the planets of the Earth group and the cores of giant planets were formed by the aggregation of solid particles (pebbles and planetozemals), which themselves formed from small dust particles, then it can be assumed with sufficient confidence that the protoplanetary disk contained a mass of dust, at least several times the mass of the Earth. It also follows from analytical and numerical models of planet formation that a significant part of the dust did not enter the planets, but was removed from the inner region of the Solar system by radiation pressure and stellar wind. As a result, this dust became part of the Oort Cloud, where, in addition to dust, ice bodies of different scales are also present in large quantities. 

One evidence of the dust presence in the Solar system is the zodiacal light. It is caused by dust in the inner region of the Solar system, where it was carried out by comets from the Oort cloud \cite{GunFulBlu20}, \cite{ShuZol22}, formed as a result of crushing asteroids in the asteroid belt \cite{Fes58,Goretal99} or came from the Kuiper belt by some means \cite{BacDasSte95,Goretal00}. Asteroid dust by its origin can be called the dust of the second generation \cite{AngMar21}. The evolution of dust fluxes created by comets in the solar system was considered in the work \cite{TutSizVer21}. The migration of dust in the Solar system and the formation of the zodiac cloud is described in the review \cite{MarIpa23}. We also consider the transfer of dust from the Oort cloud, but under the influence of of another effect -- the scattering of dust particles on the PBH. Therefore, we are primarily interested in the question of how much dust is concentrated in the inner Oort cloud. According to an estimate made in \cite{KaiVol22} based on data from observations of long-period comets, the mass of comets in the Oort cloud is $1-5$ Earth masses. As a reasonable hypothesis, it can be assumed that the total mass of non-water dust has the same order of magnitude. 

Observations in the IR range of the dust thermal radiation using the IRAS space telescope allowed one to obtain data (confirmed also in the observations of COBE and Spitzer) on dust in the zodiacal light formation area  \cite{Nesetal10}.  Dust in the Solar system can be detected using radio telescopes designed to study relic radiation. Such studies, for which dust is a source of interfering background, began to be conducted during the period of operation of the COBE telescope. Data from the observation of dust in the Solar system by a telescope ``Planck'' are given in \cite{Ver16}, however, according to available data, it is still difficult to estimate the total mass of dust in the inner Oort Cloud, since dust heating becomes weak at large distances from the Sun. 

Direct observations from the STEREO spacecraft (Solar TErrestrial RElations Observatory) give the value of the nanopowder flux (dust particles with masses of $\sim10^{-24}$~kg) at the level of $\sim 3\times10^{-6}$~$\mu$g~m$^{-2}$~year$^{-1}$ \cite{ManMeyCze14}, however, this component of cosmic dust, if formed by metallic or graphite compounds, is short-lived (quickly swept out) and therefore probably due to its origin to the fragmentation process of small bodies of the solar system. Another direct observation of dust is the counting of cosmic dust particles in the ice of Antarctica. As mentioned above, this calculation gives the value of $\sim 3.0-5.6$~$\mu$g~m$^{-2}$~year$^{-1}$ for the dust flow near the Earth \cite{Rojetal21}.

The type of distribution of dust particles by their size (mass) should have a significant effect on the amount of dust flow near the Earth's orbit. Only particles with sizes greater than 10~microns (and non-metallic non-graphite particles with sizes $\leq 0.02$~microns) do not experience a noticeable effect of radiation pressure \cite{BurLamSot79}. The function of particle distribution in the inner Oort cloud is unknown, and indirectly it can be judged only by the distribution of comet dust at shorter distances from the Sun, but this method does not provide reliable information about free dust in the inner Oort Cloud. Due to the existing uncertainties, further, the total mass of dust means only the total mass of those particles that are weakly affected by radiation pressure.

\begin{figure}
	\begin{center}
\includegraphics[angle=0,width=0.5\textwidth]{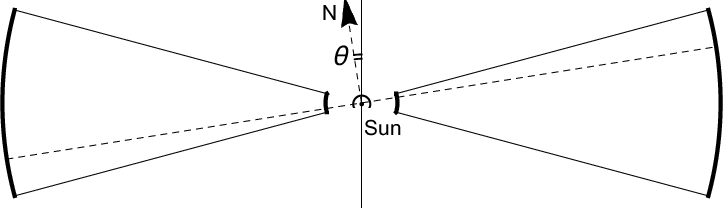}
	\end{center}
\caption{A cross section of a toroidal dust cloud. The shaded curve shows the cross section of one of the planes in which the orbits of the dust particles lie. The vector $\vec N$ is the normal to this plane, forming an angle $\theta$ with the normal to the plane of the planetary disk.}
\label{gr1}
\end{figure}

Thus, the total mass and distribution of dust in the inner Oort Cloud currently remain largely free parameters. For the calculation demonstrating the possible expected value of the dust flow, we chose one simple model, and all model assumptions are listed further in the text. As a working example, we will construct a model dust distribution that resembles a torus in configuration, but with a cross section that is bounded by two straight lines and two concentric circles with radii $R_{\rm min}$ and $R_{\rm max}$.  The cross section of the plane passing through the Sun and the normal to the planetary disk is shown in Fig.~\ref{gr1}.   

The orbits of the dust particles lie in planes passing through the Sun. Fig.~\ref{gr1} shows a cross section of one of these planes having a normal vector $\vec N$. We assume that the angle $\theta$ between $\vec N$ and the normal to the planetary disk is limited by the interval $0<\theta<\theta_{\rm max}$. That is, our distribution over orbits generally has a toroidal shape, and becomes spherically symmetric only at $\theta_{\rm max}=\pi$ (if we take into account the movement in two directions).
We further assume that the statistical distribution of orbits corresponds to the solid angles swept by the vector $\vec N$. If we introduce a spherical coordinate system with angles $\phi$ and $\theta$, then the distribution over the angles characterizing the direction of the vector $\vec N$ will take the form
\begin{equation}
P_1d\phi d\theta=-\frac{\sin\theta}{2\pi(1-\cos\theta_{\rm max})}d\phi d\theta.
\end{equation} 

Let's denote by $f(a,e)$ the distribution function of the dust particles along the semimajor axes $a$ and the eccentricities $e$ of their orbits, assuming that $\int f(a,e)dade=1$ when integrated over the entire allowable region. 
The radial distance from $r$ to $r+dr$ a dust particle passes during $dt$ twice in its orbital period $T$, so
the mass distribution of dust along the radius has the form
\begin{equation}
dF_M=P_1d\phi d\theta f(a,e)dade\frac{2dt}{Tdr}dr,
\label{distrmain}
\end{equation} 
where \cite{LL-I} 
\begin{equation}
T=\frac{\pi GM_\odot}{2^{1/2}|\varepsilon|^{3/2}},
\end{equation}
and the total energy per unit mass
\begin{equation}
\varepsilon=\frac{v^2}{2}-\frac{GM_\odot}{r}=-\frac{GM_\odot}{2a}.
\label{varr2a}
\end{equation} 
The value of $dt/dr$ is given by the well-known equation of motion in an elliptical orbit \cite{LL-I} 
\begin{equation}
\frac{dt}{dr}=\frac{1}{\sqrt{2[\varepsilon-u(r)]-l^2/r^2}},
\end{equation}
where $u=-GM_\odot/r$, and the relationship of angular momentum per unit mass $l$ with other variables has the form
\begin{equation}
e=\sqrt{1+\frac{2\varepsilon l^2}{G^2M_\odot^2}}.
\label{eeqv}
\end{equation}

\begin{figure}
	\begin{center}
\includegraphics[angle=0,width=0.45\textwidth]{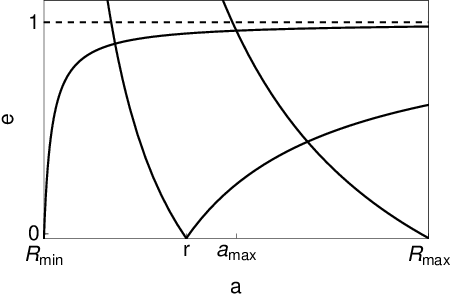}
	\end{center}
\caption{A quadrangular curved region bounded by solid curves contains acceptable distribution parameters for dust particles moving inside a toroidal distribution and passing through a point at a distance of $r$ from the Sun in the case of $r<a_{\rm max}$.}
\label{gr3}
\end{figure}

Let us now consider the conditions under which the orbit of a dust particle lies inside the toroidal distribution and passes through a point at a distance of $r$ from the Sun. These conditions have the form
\begin{equation}
R_{\rm min}\leq a(1-e)\leq r\leq a(1+e) \leq  R_{\rm max}.
\label{usl}
\end{equation}
Two external conditions define the entire domain of the distribution function definition. In Fig.~\ref{gr3} it is bounded by the axis $e=0$ and curves passing through the points $(R_{\rm min},0)$ and $(R_{\rm max},0)$. Suppose that the simplest case $f=const$ is realized, then from the condition $\int f(a,e)dade=1$, integrating over the above area, we obtain
\begin{equation}
f=\left[R_{\rm max}\ln\left(\frac{R_{\rm max}}{a_{\rm max}}\right)-R_{\rm min}\ln\left(\frac{a_{\rm max}}{R_{\rm min}}\right)\right]^{-1},
\end{equation}
where we denote $a_{\rm max}=(R_{\rm min}+R_{\rm max})/2$.

The whole condition (\ref{usl}) corresponds to a quadrangular curved area bounded by curves in Fig.~\ref{gr3} for the case of $r<a_{\rm max}$ (and a similar area for $r>a_{\rm max}$). In the numerical study of the scattering of dust particles on the PBH, it is necessary to select an array of points near the orbit of the PBH and consider the dust particles passing through these points in different directions at different speeds. If a point in space is specified, then the distribution of dust particles in orbits has a parameter range in the specified curved quadrangular region, as well as various variable parameters $\phi$ and $\theta$ for the inclination of the orbit (the direction of the normal vector). 

\begin{figure}
	\begin{center}
\includegraphics[angle=0,width=0.45\textwidth]{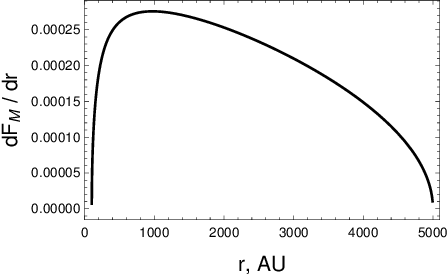}
	\end{center}
\caption{Model mass distribution function of dust particles along the radius in the inner Oort cloud.}
\label{gr4}
\end{figure}

The expression $(1/T)dt/dr$ can be converted to the following convenient form
\begin{equation}
\frac{1}{T}\frac{dt}{dr}=\frac{r^{1/2}}{2\pi a^{3/2}}\frac{1}{\left[e^2-\left(1-r/a\right)^2\right]^{1/2}}.
\end{equation}
If we integrate the distribution over all $\phi$, $\theta$, $e$, $a$, then we can get a mass distribution of dust particles over the radius of $dF_M/dr$. Integration over $\phi$, $\theta$ is trivial, and numerical integration over quadrangular curved regions is performed by dividing the integral into three parts for each of the cases $r<a_{\rm max}$ and $r>a_{\rm max}$. In double integrals, the internal integration of $e$ is performed analytically, and the external integration of $a$ is performed numerically. As a result, we get the distribution of $dF_M/dr$, shown in Fig.~\ref{gr4}. For example, $R_{\rm min}=100$~AU and $R_{\rm max}=5000$~AU is assumed  (this interval contains the PBH). To calculate the density, it is necessary to divide $dF_M/dr$ by $4\pi r^2$.
Integrating $dF_M/dr$ by $r$ from $R_{\rm min}$ to $R_{\rm max}$, up to computational errors, we get $\int (dF_M/dr)dr=1$.


\section{Dust particle scattering: numerical calculation}
\label{algsec}

Suppose that the orbit of the PBH lies in a plane characterized by the normal vector $\vec N_0$, and has the orbit parameters $a_0$ and $e_0$. The orbit is oriented on the plane in a certain way, and the planet has some initial angle of orbital motion. We select the average parameters of the PBH orbit in accordance with the works \cite{Batetal19,BatBro21}, but consider several different options for parameters that differ from the average.

Let's introduce two Cartesian rectangular coordinate systems centred in the Sun: 1. the barycentric coordinate system $(x,y,z)$, in which the z axis is directed along the axis of the planetary disk, and 2. the coordinate system $(x',y',z')$, in which the axes $x'$ and $y'$ lie in the plane of the PBH orbit. One coordinate system can be translated into another by rotating by an angle $\phi$ in the plane $x,y$ and by an angle $\theta$ relative to the axis $y'$. The transformation of the radius vector components in this case has the form 
\begin{eqnarray}
x&=&x'\cos\theta\cos\phi-y'\sin\phi+z'\sin\theta,
\nonumber
\\
y&=&x'\cos\theta\sin\phi+y'\cos\phi+z'\sin\theta\sin\theta,
\nonumber
\\
z&=&-x'\sin\theta+z'\cos\theta.
\label{coordtransf}
\end{eqnarray}
In the coordinates $(x',y',z')$ of the equation of motion of the PBH \cite{LL-I} 
\begin{eqnarray}
x'&=&a(\cos\xi-e),
\nonumber
\\
y'&=&a\sqrt{1-e^2}\sin\xi,
\nonumber
\\
z'&=&0,
\label{eqxyz9pl}
\end{eqnarray}
where the parameter $\xi$ varies from 0 to $2\pi$ over the orbital period. The variable $\xi$ and the time $t$ are connected by the Euler equation, but we do not need to solve it. In the process of numerical calculation, the specified interval $\xi$ is divided into many steps, and each increment of $\Delta\xi$ corresponds to an increment
\begin{equation}
\Delta t=\frac{dt}{d\xi}\Delta\xi=\left(\frac{a^3}{GM_\odot}\right)^{1/2}(1-e\cos\xi)\Delta\xi.
\label{dtdxi}
\end{equation}
The numerical algorithm summarizes the dust flows into the Earth's orbit produced by the PBH as it moves along the orbit, taking into account the duration $\Delta t$ of each step. The speed of the PBH $\vec V$ in the system $(x,y,z)$ is found by differentiating the equations (\ref{eqxyz9pl}) in time  and converting the results with the formulas (\ref{coordtransf}).

Let's assume that the PBH is at some point with the radius vector $\vec r=(x,y,z)$. Consider a dust particle moving near this point at a speed of $\vec v$ in an orbit with parameters $a$, $e$. 
The velocity of the dust particle $\vec v$ forms an angle $\gamma$ with $\vec r$:
\begin{equation}
\sin\gamma=\frac{(1-e^2)^{1/2}}{2^{1/2}\left(\frac{r}{a}\right)^{1/2}\left(1-\frac{r}{2a}\right)^{1/2}}.
\label{singam}
\end{equation}
For each $r$, there are generally two angles $\gamma$, except for the points $r\to R_{\rm min}$ and $r\to R_{\rm max}$, where always $\gamma\to\pi/2$. Knowing $a$, $e$ and $r$, it is easy to find the absolute value of the velocity $v$, and then using (\ref{singam}) one can find the velocity component $\vec v_r$ along $\vec r$ and the component $\vec v_{\rm tr}$ perpendicular to $\vec r$. 

The structure of the dust cloud in the outer region of the solar system remains unknown. In the numerical calculation, we will use the dust distribution (\ref{distrmain}) in the spherically symmetric case when $\theta_{\rm max}=\pi$. This will give us a conservative estimate, because in the case of a dust cloud more concentrated towards the plane of the planetary disk, the dust density near the orbit of the PBH will be higher (with the same total dust mass), and, accordingly, the generated dust flow near the Earth will be greater. In the case of a spherically symmetric dust distribution, the perpendicular component $\vec v_{\rm tr}$ has equally probable directions when rotating relative to $\vec r$. To define the perpendicular component, we introduce a basis of vectors
\begin{equation}
\vec f_1=\frac{\vec r\times\vec k}{|\vec r\times\vec k|}, \quad \vec f_2=\frac{\vec f_1\times\vec r}{|\vec f_1\times\vec r|},
\end{equation}
where $\vec k=(0,0,1)$ is a unit vector along the direction of the z axis. Then the speed of a dust particle is decomposed as follows
\begin{equation}
\vec v=\vec v_r+v_{\rm tr}\left(\vec f_1\cos\omega+\vec f_1\sin\omega\right),
\end{equation}
where the angle $\omega$ takes random values and is played using the Monte Carlo method.

We will consider the scattering of dust particles by the gravitational field of the PBH in the approximation of a gravitational maneuver, when during the scattering the gravitational field of the Sun can be ignored, since the scattering leading to the pointing of the dust particles into the region of the Earth's orbit occurs only at a sufficiently small distance from the PBH.  Let's switch to the PBH rest system. In this system, a dust particle has a velocity $\vec w=\vec v -\vec V$ and some impact parameter $b$. Let's introduce a vector $\vec b$ with a length of $b$ and a direction from the PBH to the dust particle at the moment of minimal approaching in the case that there was no scattering. The vector $\vec b$ can be set by changing the parameters of the orbit of the dust particle $a$, $e$, $\omega$ in some small intervals, but in numerical calculation, a more convenient method is to fix the average values of $a$, $e$, $\omega$ and directly vary the magnitude and direction of $\vec b$. To this end, we introduce two unit vectors perpendicular to $\vec w$ and each other
\begin{equation}
\vec e_1=\frac{\vec r\times\vec w}{|\vec r\times\vec w|}, \quad \vec e_2=\frac{\vec e_1\times\vec w}{|\vec e_1\times\vec w|}.
\end{equation}
The components $\vec b$ in this basis are given as $b_1=b\cos\beta$, $b_2=b\sin\beta$, where the angle interval $\beta$ and the value $b$ are divided into many small intervals with further summation of the dust particle flow over all the elements of the grid.

The impact parameter is related to the scattering angle by the following equation \cite{LL-I}
\begin{equation}
b=\pm\frac{GM_{\rm PBH}}{w^2}\cot\frac{\chi}{2}.
\end{equation}
The speed of the dust particle after scattering is
\begin{equation}
\vec w'=\vec w\cos\chi-\frac{\vec b}{b}w\sin\chi,
\end{equation}
and the velocity of the dust particle after scattering in the system $(x,y,z)$ is
\begin{equation}
\vec v'=\vec w'+\vec V.
\end{equation}
Knowing the radial distance $r$ and the velocity $v'$, using the formulas (\ref{varr2a}) and (\ref{eeqv}) we calculate the final parameters of the orbit of the dust particle $a'$, $e'$. If $a'(1-e')<R_{\rm E}$, where $R_{\rm E}$ is the radius of the Earth's orbit, then we conclude that the dust particle falls into the region of the Earth's orbit and contributes to the dust flow near the Earth. In the process of numerical simulation, the sum is calculated for one orbital period of the PBH 
\begin{equation}
J=\sum_\xi\sum_i\sum_j\rho_i w_i b_j(\Delta b)_j(\Delta\beta)_j\Delta t_\xi/T_{ij},
\label{sumij}
\end{equation}
where the terms $i$ correspond to dust subsystems initially moving at different speeds $\vec v_i$ and played by the Monte Carlo method, and the terms $j$ correspond to summation in magnitude and direction of the impact parameter. The time interval $\Delta t$ corresponds to the interval of the parameter change $\xi$ according to (\ref{dtdxi}). The orbital periods $T_{ij}$ of dust particles at new orbits, which must be taken into account to calculate the dust flow near the Earth, are expressed in terms of the final major semi-axis $a'$. In the sum (\ref{sumij}), we leave only those terms for which $a'(1-e')<r_{\rm E}$. Summing up the $\xi$ intervals over the orbital period, we find the total dust flow over the orbital period, and then the total flow that was created during the existence of the Solar system. To recalculate the created dust flow to the flow on Earth, it is necessary to take into account the fraction of time that dust particles spend inside the Earth's orbit during their orbital period and the characteristic velocity of dust particles in this area. 

For the described calculation, it is necessary to find the probability of the dust particle ejecting from orbit due to the variable gravitational field of the Sun-Jupiter system (other planets make a smaller contribution), or, in other words, the lifetime of the dust particle in orbit $t_{\rm ej}$, which is the topic of the next section.


\section{Time of ejection from orbit}
\label{tejsec}

Let's consider a dust particle that, after scattering on the PBH, moved into an orbit entering the Earth's orbit, i.e. $a'(1-e')<R_{\rm E}$. In the inner region of the orbit, this dust particle will be exposed to the gravitational fields of planets, primarily Jupiter, and its orbit will begin to evolve. Since close approaches to Jupiter are unlikely, the evolution of the orbit will occur in small steps with each passage of the dust particle through the central region of the Solar system. And, since Jupiter almost accidentally turns out to be on different sides of the trajectory of the dust particle during each flyby, it exerts the effects of different signs, and the process of orbital evolution will proceed in a diffusive manner. An accurate numerical calculation of this process requires large computational resources, so in this paper we will limit ourselves to an analytical estimate in the diffusion momentum approximation with an adiabatic correction.

The characteristic distance of the dust particle from Jupiter $r$, when it is affected, has the order of the radius of the Jupiter orbit $r\sim R_J$, and the characteristic time of movement from the Jupiter orbit to the Earth orbit and back is $\Delta t\sim 2R_J/(GM_J/R_J)^{1/2}=T_J/\pi$, where $T_J$ is the orbital period of Jupiter. The dust particle with a mass $m$ is affected by Jupiter with a force of $F=GM_Jm/r^2$. Since the velocity of the dust particle during the passage of Jupiter's orbit is on average $\sqrt{2}$ times greater than the orbital velocity of Jupiter, when using the momentum approximation, it is necessary to take into account the adiabatic correction. The correction in the Weinberg form \cite{GneOst99} has the value $A=(1+(\sqrt{2})^2)^{-3/2}\simeq0.2$. The impact of Jupiter during the time $\Delta t$ will lead to a change in the specific angular momentum of the dust particle by an amount
\begin{equation}
\Delta l_1=2^{1/2}(GM_J)^{1/2}R_J^{1/2}A.
\label{l1eq}
\end{equation}
In the diffusion approximation, the change in the square of the specific angular momentum over time $t$ is
\begin{equation}
(\Delta l)^2=\sum(\Delta l_1)^2\simeq (\Delta l_1)^2 \frac{t}{2T},
\label{dell}
\end{equation}
where the summation goes for each flyby of the dust particle through the inner region of the Solar system, and it is taken into account that only about half of the time Jupiter is in that part of its orbit where it can influence the angular momentum of the dust particle. 

Let us now take into account that the dust particles in question move in very elongated orbits with $1-e'\ll 1$. In this case, one can write
\begin{equation}
e'=\left(1-\frac{l^2}{GM_\odot a'}\right)^{1/2}\simeq1-\frac{l^2}{2GM_\odot a'},
\end{equation}
and the minimum distance from the Sun 
\begin{equation}
r_{\rm min}=a'(1-e')\simeq\frac{l^2}{2GM_\odot}
\label{rminappr}
\end{equation}
depends only on $l$. In order for the dust particle to leave the region of the Earth's orbit, a characteristic change of $\Delta r_{\rm min}\sim r_{\rm min}$ is necessary, which corresponds to $\Delta l\sim l/2$. Substituting in (\ref{dell}) and expressing $l$ from (\ref{rminappr}) under the condition $r_{\rm min}\simeq R_{\rm E}$, we get the characteristic time for which the orbit of the dust particle will leave the region of the Earth's orbit,
\begin{equation}
t_{\rm ej}\simeq\frac{T}{2A^2}\frac{M_\odot}{M_J}\frac{R_E}{R_{\rm J}}.
\end{equation}
Numerically 
\begin{equation}
t_{\rm ej}\simeq2.7\times10^6\left(\frac{T}{10^3\mbox{~years}}\right)\mbox{~years}.
\end{equation}
In the cases we are considering, the time $t_{\rm ej}$ is always less than the lifetime of the Solar system $t_s=5$~billion years, but in general, the summation of the dust flow for each parameter region must be performed in time
$t=min\left\{t_s,t_{\rm ej}\right\}$.

To verify the estimate (\ref{l1eq}), we performed a numerical experiment of $10^3$ simulations of the passage of the dust particle through the inner region of the Solar system. Each simulation began with a configuration when the dust particle was 15~AU from the Sun, and the initial position of Jupiter in orbit was set by a random number generator: the phase of its movement was played out randomly in the range from 0 to $2\pi$. The semimajor axis of the dust particle orbit was assumed to be 500~AU, and its eccentricity is $0.998$ -- in this case, the minimum approach of the dust particle with the Sun is 0.5~AU, i.e. the dust particle goes inside the Earth's orbit. Further, the motion of the dust particle and Jupiter was traced along their elliptical undisturbed orbits (the change in angular momentum in this estimate is considered as a correction of the next order of smallness). For each position of the dust particle and Jupiter in the orbits, the force vector was calculated, and as they moved, the increment of the specific angular momentum was calculated
\begin{equation}
\Delta \vec l_1=\int dt[\vec r\times \vec F]/m.
\end{equation}

\begin{figure}
	\begin{center}
\includegraphics[angle=0,width=0.45\textwidth]{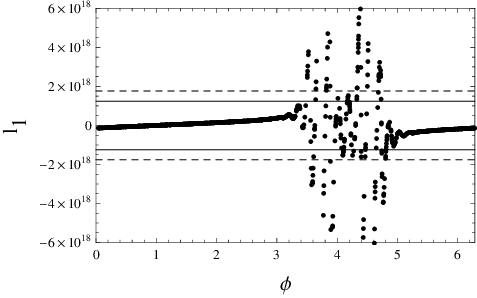}
	\end{center}
\caption{The increment of the specific angular momentum $l_1$ (in units of cm$^2$~s$^{-1}$) during one flyby through the inner region of the Solar system, depending on the initial phase of Jupiter's position on the orbit. The dots represent the results of numerical simulations, the solid curve shows the value (\ref{l1eq}) (multiplied by $\pm1$), and the shaded curve shows the average value of the modulus $l_1$ in a numerical experiment (multiplied by $\pm1$). }
	\label{grexp}
\end{figure}

The result of the experiment is shown in Fig.~\ref{grexp}. It can be seen that, depending on the position of Jupiter in orbit, there are regions of large and small angular momentum increase. For about 2/3 of the interval of the initial phases of Jupiter's orbit, the momentum increase of the dust particle is small, and the main part of the set occurs for about 1/3 of the initial positions of Jupiter. For the average increment of $l_1$, good agreement was obtained with (\ref{l1eq}). The numerical value of the average increment is only 1.4 times higher than the value (\ref{l1eq}), and the average ejection time, respectively, is only 1.2 times greater, which is close to the accuracy of our numerical simulation. 

Note also that for sufficiently large particles, the light pressure and the Poynting-Robertson effect are not important, so these effects lead to the removal of only very fine dust particle from orbits passing through the Earth's orbit region. In this paper, we consider only those sufficiently massive dust particles (with dimensions of $\geq10$~microns) that are swept out only due to the gravitational influence of Jupiter. Consideration of a more general task requires consideration of additional factors and is beyond the scope of this work.


\section{Results}
\label{resultsec}

\begin{figure}
	\begin{center}
\includegraphics[angle=0,width=0.5\textwidth]{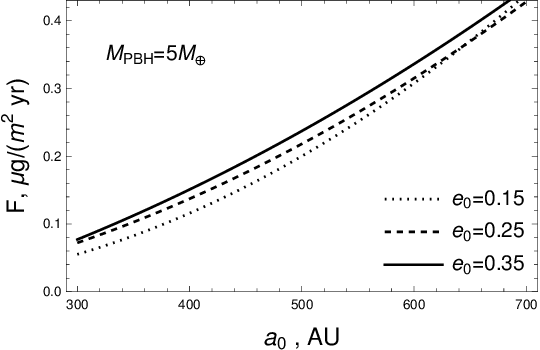}
	\end{center}
\caption{The dust flow near the Earth's orbit, depending on the large semi-axis of the PBH (9th planet) orbit in the case of its mass $M_{\rm PBH}=5M_\oplus$ for the eccentricities of the orbit $e_0=0.15$, $0.25$ and $0.35$ (dotted, dashed and solid curves, respectively).  }
	\label{gr5}
\end{figure}

\begin{figure}
	\begin{center}
\includegraphics[angle=0,width=0.5\textwidth]{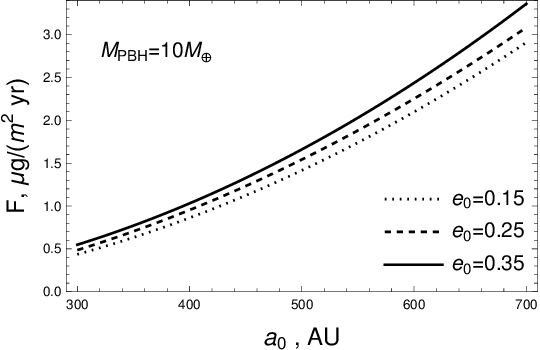}
	\end{center}
\caption{The same, but for $M_{\rm PBH}=10M_\oplus$.}
\label{gr10}
\end{figure}

In the numerical calculation, for example, we used the spherically symmetric distribution of dust according to orbital parameters, described in the Section~\ref{tor} with $R_{\min}=100$~AU, $R_{\max}=5000$~AU. The total mass of dust in this cloud $M_d$ entered the result (dust flow near the Earth) as a common multiplier, therefore, we present the results for the case of $M_d=5$ Earth masses, and for other values of $M_d$ the result can be obtained by simple recalculation. The calculation used the algorithm described in Section~\ref{algsec} and the time of dust particles ejection from orbit found in Section~\ref{tejsec}. The calculation results are shown in Fig.~\ref{gr5} and \ref{gr10}. Calculations were performed for three values $a_0=300$, 500 and 700~AU, and the intermediate points of the curves represent a parabolic interpolation.

As can be seen from the graphs, in the case of $M_d=5M_\oplus$, and PBH with a mass of $5M_\oplus$ with the specified orbital parameters can create a dust flow near the Earth of $\sim0.05-0.4$~$\mu$g~m$^{-2}$~year$^{-1}$, and PBH with a mass of $10M_\oplus$ with the specified orbital parameters can create a dust stream near the Earth of $\sim0.5-3$~$\mu$g~m$^{-2}$~year$^{-1}$, which is comparable to the actually observed flow of \cite{Rojetal21}. Thus, a significant part of the dust near the Earth's orbit could have been directed to this area from the inner Oort Cloud during scattering on the PBH (or on the 9th planet). The dependence on $M_{\rm PBH}$ is quite strong, and in the case of $M_{\rm PBH}=1M_\oplus$, the flow is several times smaller (in this case, computational errors become large in our calculations, and we do not provide the results). The low accuracy at low mass $M_{\rm PBH}$ is due to the fact that in this case the scattering is weak, so a more frequent grid is required according to the aiming parameter $b$. In the case $M_d=5M_\oplus$, the difference of the curves in Fig.~\ref{gr5} is comparable to the computational error, but for $M_d=10M_\oplus$ the calculation accuracy is higher, and in Fig.~\ref{gr10} a slight increase in the dust flow is visible with an increase in the eccentricity $e_0$ of the PBH orbit. 

Despite the fact that when moving away from the Sun from distances of 300~AU to 700~AU, the dust density in our dust cloud model decreases, the dust flow on Earth increases with an increase in the large semi-axis of the orbit of the PBH within these limits. We explain this behaviour by the fact that the orbital velocities of the dust particles are lower at large distances, so they can more easily lose the transverse velocity component when scattered by the PBH and begin to fall towards the Earth.    

Dust particles arriving from the periphery of the Solar system have an average velocity $\sqrt{2}$ times higher than dust particles constantly orbiting the Sun in near-Earth orbits. Thus, these dust particles represent a high-speed component. 

In the described calculation, a model spherically symmetric dust distribution was used. It should be noted that if the dust cloud is compressed to the plane of the ecliptic (as shown in Fig.~\ref{gr1}), then with the same total mass of dust in the cloud, it will have a higher density, and the dust flow near the Earth will be several times greater than what was obtained in our calculation.

In this paper, we are talking only about dust particles. It is possible that in the region of 300-700~AU from the Sun there are cometary nuclei with sizes of hundreds of meters to kilometers. The question of the rate of their injection into the inner region of the Solar System when interacting with the PBH requires a separate study.


\section{Conclusion}
\label{zaklsec}

With a fairly conservative choice of dust cloud parameters (masses of $\sim 5M_\oplus$ within $\sim5000$~AU), the dust flow produced by the PBH during gravitational scattering on Earth can have a value of $\sim0.1-3$~$\mu$g~m$^{-2}$~year$^{-1}$, which is close to the dust flow actually observed near the Earth $\sim 3.0-5.6$~$\mu$g~m$^{-2}$~year$^{-1}$ \cite{Rojetal21}. If the PBH has a lower mass, or if the total mass of dust is less than expected in the calculation, then the dust flow will also be correspondingly less. 

It is generally believed that the observed dust is brought to Earth by passing comets or comes from collisions and destruction of objects in the asteroid belt. We see that the presence of the captured PBH or planet on the periphery of the Solar System can create a dust flow of the same order of magnitude in the presence of a significant amount of it in the inner Oort cloud. The uncertainty in our calculations is related to strong uncertainties in the mass and distribution of dust at distances of $\sim300-700$~AU from the Sun. Currently, only a few large objects from this area are known, and only very approximate estimates can be obtained with an accuracy of the order of magnitude. In the future, when the structure, mass and composition of this cloud are clarified, it will be possible to make more reliable predictions about the flow of dust into the inner regions of the Solar System and about the presence or absence of a captured PBH or planet. 

If the calculated dust flow produced by the PBH turned out to be greater than the observed flow, then this would eliminate the dust cloud models and limit the mass of the PBH. Even a small contribution to the total flow of the considered high-speed dust with a chemical composition different from that of the dust from the asteroid belt may be of interest for understanding the structure of the inner Oort Cloud. High-speed dust near the Earth's orbit may also play a role in the degradation of the functioning of optical and other sensors on spacecraft. 

Of course, not all of the observed dust comes from beyond the orbit of Neptune. The dust component has several subsystems (comet dust, dust from asteroid crushing, etc.). The new subsystem considered in this article can produce only a small part of the total dust flow. The described calculation showed that, with fairly plausible assumptions, the dust flow from outside the Neptune orbit may be at the level of the observed flow, however, it cannot be argued that this trans-Neptune dust really prevails in this flow. It can be only a very small part, 10\% or much less. The exact value, as explained above, depends on the mass and configuration of the dust cloud and the mass of the distant planet. 

It would be possible to consider a flow of dust not from a planet captured in the Solar System or an PBH, but from objects flying through the Solar System from the halo of the Galaxy at a speed of $\sim200$~km~s$^{-1}$. The calculation, however, shows that the dust flow from the gravitational influence of passing bodies of planetary mass is several orders of magnitude smaller. Thus, based on the flow of dust, it is impossible to limit the number of PBHs or free-flying planets in the Galaxy. The probability of capture into the Solar system of a planet or a PBH with a mass of about 5-10 Earth masses for the entire lifetime of the Solar system is very small, $\sim10^{-4}-10^{-3}$ \cite{LinLoe18}. But if such an object was nevertheless captured in the Solar System, then, as we have shown, it is capable of creating a significant dust flow on Earth, comparable to what is actually observed.

The authors are grateful to Reviewers of the article for useful comments. This work is supported by the Russian Science Foundation grant 23-22-00013,

https://rscf.ru/en/project/23-22-00013/.


\section*{Appendix. Simple estimates}

To study the scattering of dust particles by the gravitational field of the PBH, we performed a numerical calculation. In this section, we will show how the dust flow can be estimated in order of magnitude in a simple way. If the radius of the dust cloud is equal to $r_i$, then its density  
\begin{equation}
\rho_d\simeq\frac{3M_d}{4\pi r_i^3}.
\end{equation}
We look at the dust particle moving at the orbit around the Sun in the mean distance $r_i$ at the speed $v_i\simeq(GM_\odot/r_i)^{1/2}$, and let past the dust particle on the impact parameter $b$ near the PBH with a mass of $M_{\rm PBH}$. We suppose that the initial orbits of both the dust particle and the PBH are not circular, not very elongated, and $1-e\sim1$. In this case the relative velocity of the dust particle and the PBH is $v_{\rm rel}\sim v_i$. During flight by, the dust particle receives a speed boost towards the point of maximum approach
\begin{equation}
\Delta v\simeq \frac{2GM_{\rm PBH}}{bv_{\rm rel}}.
\label{dvb}
\end{equation}
This expression is the limiting case of momentum approximation, which is valid here by order of magnitude. The speed of the dust particle after the interaction is denoted through $v_f$. If the speed boost significantly compensates for the initial velocity, then the dust particle will lose angular momentum and begin to fall almost towards the Sun in an elongated orbit. The condition that the dust particle will enter the Earth's orbit when approaching the Sun can be found from the condition of conservation of angular momentum
\begin{equation}
l\sim r_iv_f\sim v_{\rm E} R_{\rm E},
\label{lcond}
\end{equation}
where $R_{\rm E}$=1~AU is the radius of the Earth's orbit, and the velocity is $v_{\rm E}\simeq(GM_\odot/R_{\rm E})^{1/2}$. From (\ref{lcond}) we get
\begin{equation}
v_f\sim v_i\left(\frac{R_{\rm E}}{r_i}\right)^{1/2}\ll v_i.
\label{vfviee}
\end{equation}
The initial velocity is $\Delta v\sim v_i$ therefore one get 
\begin{equation}
b\simeq \frac{2GM_{\rm PBH}}{v_i^2}.
\label{bvi2}
\end{equation}
For typical parameters, the value of $b$ is $\sim 3$~million km. As follows from (\ref{dvb}), the final velocity of the dust particle is $v_f$ for a certain narrow interval $\delta b$ of the impact parameters
\begin{equation}
\frac{\delta b}{b}\simeq \frac{v_f}{v_i}.
\end{equation}

If the initial velocity of the dust particle was directed outside the plane perpendicular to the trajectory of the PBH, then after compensating the velocity
$v_i$ the dust particle will gain additional velocity perpendicular to the radius vector directed away from the
Sun. For the dust particle to still be able to enter within Earth’s orbit, this additional velocity should not
exceed $v_f$. It follows that only dust particles with initial velocities within an angle $\phi\simeq v_f/v_i$ relative to the
specified plane can enter inside Earth’s orbit. On the contrary, the gain in velocity along the radius vector
will not affect the entry into Earth’s orbit and will not change the order of magnitude.

Putting together all the above, we get that the mass of dust directed into the Earth's orbit per unit of time during scattering by the PBH is
\begin{equation}
\frac{dM_d}{dt}\simeq b(\delta b) \phi\rho_dv_i.
\end{equation}

Let us denote by $T$ the full orbital period of the dust particle after scattering on the PBH and the time it spends inside Earth’s
orbit as $\Delta t$. The probability of a dust particle being inside Earth’s orbit can be estimated as follows
\begin{equation}
P_t=\frac{\Delta t}{T}\sim\frac{(2R_{\rm E}/v_{\rm E})}{(2r_i/v_i)}.
\end{equation}
Taking into account the time of ejection of the dust particle $t_{\rm ej}$, the stationary density will be
\begin{equation}
\rho_s\sim\frac{P_t(dM_d/dt)t_{\rm ej}}{(4\pi R_{\rm E}^3/3)},
\end{equation}
and the flow
\begin{equation}
F=\rho_sv_{\rm E}\simeq\frac{9}{4\pi^2}\frac{GM_{\rm PBH}^2M_dt_{\rm ej}}{M_\odot R_{\rm E} r_i^4}.
\label{total}
\end{equation}
Numerically one gets
\begin{eqnarray}
F&\simeq&0.4\left(\frac{M_{\rm PBH}}{10M_\oplus}\right)^2
\left(\frac{M_d}{5M_\oplus}\right)
\left(\frac{t_{\rm ej}}{2.5\times10^6\mbox{~years}}\right)
\nonumber
\\
&\times&
\left(\frac{r_i}{500\mbox{~AU}}\right)^{-4}
\mbox{~$\mu$g~m$^{-2}$~year$^{-1}$}.
\label{chisl}
\end{eqnarray}


\begin{thebibliography}{99}

\bibitem{Batetal19} K. Batygin, F. C. Adams, M. E. Brown, and J. C. Becker, Phys. Rep. {\bf 805}, 1 (2019).

\bibitem{SchUnw20} J. Scholtz and J. Unwin, Phys. Rev. Lett. {\bf 125}, 051103 (2020).

\bibitem{Napetal21} K. J. Napier et al., Planetary Science Journal {\bf 2}, 59 (2021).

\bibitem{LykIto23} P. S. Lykawka, and T. Ito, The Astronomical Journal {\bf 166} 118 (2023). 

\bibitem{ZelNov66} Ya.B. Zel’dovich, and I.D. Novikov, Sov. Astron. {\bf 10}, 602 (1967).

\bibitem{Dol18}  A. D. Dolgov, Phys. Usp. {\bf 61}, 115 (2018).

\bibitem{CarSenYok20} B. Carr, Y. Sendouda, and J. Yokoyama, Reports on Progress in Physics {\bf 84}, 116902 (2021).

\bibitem{Blietal16} S. Blinnikov, A. Dolgov, N. K. Porayko, and K. Postnov, JCAP {\bf 11}, 036 (2016).

\bibitem{DolPos20} A. Dolgov and K. Postnov, JCAP {\bf 07}, 063 (2020).

\bibitem{Guoetal23} S.-Y. Guo, M. Khlopov, X. Liu, L. Wu, Y. Wu, and B. Zhu, arXiv:2306.17022 [hep-ph].

\bibitem{IvaNasNov94} P. Ivanov, P. Naselsky, and I. Novikov, Phys. Rev. D. {\bf 50},  7173 (1994).

\bibitem{CarKuh20} B. Carr  and F. Kuhnel, Annual Review of Nuclear and Particle Science {\bf 70}, 355 (2020).

\bibitem{GreKav20} A. M. Green and B. J. Kavanagh, J. Phys. G {\bf 48}, 043001 (2021).

\bibitem{Rojetal21} J. Rojas et al., Earth and Planetary Science Letters {\bf 560}, 116794 (2021). 

\bibitem{BatBro21}  K. Batygin and M. E. Brown, Astrophys. J. Lett., {\bf 910}, L20 (2021).

\bibitem{SirLoe21} A. Siraj and A. Loeb, arXiv:2103.04995 [astro-ph.CO].

\bibitem{SirLoe21-2}  A. Siraj and A. Loeb, Research Notes of the AAS {\bf 5}, 145 (2021).

\bibitem{Luoetal12}  Y. Luo et al.,  Astrophys. J. {\bf 751} 16 (2012).

\bibitem{DokVol}L. V. Volkova, and V. I. Dokuchaev, Journal of Experimental and Theoretical Physics Letters {\bf 60}, 76 (1994). 

\bibitem{SirLoe20} A. Siraj and A. Loeb, Astrophys. J. Lett. {\bf 898}, L4 (2020).

\bibitem{ArbAuf20} A. Arbey  and J. Auffinger, arXiv:2006.02944 [gr-qc].

\bibitem{BurLamSot79} J. A. Burns, P. L. Lamy, and S. Soter, Icarus {\bf 40} 1 (1979). 

\bibitem{GunFulBlu20} B. Gundlach, M. Fulle, and J. Blum, Monthly Notices of the Royal Astronomical Society {\bf 493}, 3690 (2020).

\bibitem{ShuZol22} B. M. Shustov, R. V. Zolotarev, Astronomy Reports {\bf 66}, 179 (2022).

\bibitem{Fes58} V. G. Fesenkov, Soviet Astronomy {\bf 2}, 303 (1958).

\bibitem{Goretal99} N. Gorkavyi, L. Ozernoy, J. Mather, and T. Taidakova, ASP Conference Series {\bf 207}, 462 (1999); arXiv:astro-ph/9910551

\bibitem{BacDasSte95} D. E. Backman, A. Dasgupta, and R. E. Stencel,  Astrophys. J. Lett. {\bf 450}, L35 (1995).

\bibitem{Goretal00} N. N. Gorkavyi, L. M. Ozernoy, T. Taidakova, and J. C. Mather, arXiv:astro-ph/0006435.

\bibitem{AngMar21} G. D'Angelo  and F. Marzari, Monthly Notices of the Royal Astronomical Society {\bf 509}, 3181 (2022).

\bibitem{TutSizVer21}A. V. Tutukov, M. D. Sizova, and S. V. Vereshchagin, Astronomy Reports {\bf 65}, 884 (2021).

\bibitem{MarIpa23} M. Ya. Marov, and S. I. Ipatov, Phys. Usp. {\bf 66} 2 (2023).

\bibitem{KaiVol22} N. A. Kaib and K. Volk, Chapter in press for the book Comets III, edited by K. Meech and M. Combi, University of Arizona Press; arXiv:2206.00010 [astro-ph.EP].

\bibitem{Nesetal10} D. Nesvorný, P. Jenniskens, H. F. Levison, W. F. Bottke, D. Vokrouhlický, and M. Gounelle, Astrophys. J. {\bf 713}, 816 (2010).

\bibitem{Ver16} O. V. Verkhodanov, Phys. Usp. {\bf 59} 3 (2016).

\bibitem{ManMeyCze14} I. Mann, N. Meyer-Vernet, and A. Czechowski, Physics Reports {\bf 536}, 1 (2014).

\bibitem{LL-I} L. D. Landau and E. M. Lifshitz, Mechanics: Course of Theoretical Physics, Vol. 1 (Butterworth-Heinemann, 1976).

\bibitem{GneOst99} O. Y. Gnedin and J. P. Ostriker, Astrophys. J. 513, 626
(1999).

\bibitem{LinLoe18} M. Lingam, and A. Loeb, Astron. J. {\bf 156}, 193 (2018).

\end{thebibliography}
\end{document}